%Paper: hep-th/9304061
%From: jds@math.upenn.edu
%Date: Thu, 15 Apr 93 16:07:51 -0400

\input amstex
\NoBlackBoxes
 \magnification=\magstep1
\def\pp{\vskip1ex}
\def\pc{\{ {\it Post conference:}}
\def\newcommand{\def}
\def\renewcommand{\def}

\def\R{{\Bbb R}}
\def\C{{\Bbb C}}
\def\CC{\Cal C}
\def\config{F({\Bbb R}^2,k)}

\def\Section#1{{\bf #1}}
\def\cite#1{[{\bf #1}]}

\renewcommand\a{{\cal  O}}

\newcommand\SS{{\Bbb S}}
\newcommand\bull{{\bullet}}

\newcommand\Cat{{\cal C}}
\newcommand\1{{1\!\!1}}

\newcommand\BC{{\roman  B}}
\newcommand\op{{\text{op}}}

\newcommand\OP{{\cal O}}
\newcommand\BB{{\Bbb B}}

\newcommand\Ainf{$\text{A}_\infty$}

\newcommand\part{\partial}
\newcommand\Om{\Omega}

\renewcommand\Cat{{\cal C}}

\newcommand\Cinf{C^\infty}

\def\p{\phi}
\def\Bbb{\pmb}
\def\bull{\bullet}
\def\Om{\Omega}
\def\Bbb{\pmb}
\def\compose{\circ}
\def\righthookarrow{\hookrightarrow}
\def\integral{\int}

\def\cal{\Cal}
\def\16N{d\lbrack\phi_1,\dots,\phi_N\rbrack+ \Sigma_1^N\pm \lbrack\phi_1,\dots,
d\phi_i,\dots,\phi_N\rbrack= \Sigma^{N-1}_{Q=2}\pm
\lbrack\lbrack\phi_{i_1},\dots,\phi_{i_Q}\rbrack,\phi_{i_{Q+1}},\dots,\phi_{i_N}\rbrack}
 \hsize=15cm
 \vsize=19.5cm
\def\parag{\vskip3ex}
 \pageno=1
 \headline={\ifnum\pageno>1 \tenbf
 Closed SFT and operads      \hss Stasheff\else\hfil\fi}
\line {\hss UNC-MATH-93/1}
 \line {\hss April  15, 1993}
\line {\hss hep-th/9304061 }

 \vskip 5ex\noindent
\centerline{\bf Closed string field theory, strong homotopy Lie algebras}
\parag
\centerline{\bf and the operad actions of moduli spaces}
\parag
\parag

\centerline{Talk given at the Conference on Topics in Geometry and Physics}

\centerline{USC - November 6, 1992}

 \vskip 5ex\noindent
 \centerline{Jim Stasheff}
\plainfootnote*{Supported in part by NSF grant  DMS-9237029.}
%\address Math Dept, UNC, Chapel Hill NC 27599-3250 \endaddress
%\email jds\@math.unc.edu \endemail

 \centerline{jds\@math.unc.edu}
 \vskip 5ex

I'd like to thank Bob Penner for organizing such a well-coordinated
conference.  `A funny thing happened on the way' to this conference.
Between the time that Bob set it up initially and last Monday (November
2, 1992), there has
been a flood of new results.  Being on leave at the University of
Pennsylvania, I was able to learn first hand of Y.-Z. Huang's success in
following my suggestion that something like an `operad' was present in his
approach to VOA's (vertex operator algebras).  Gregg Zuckerman invited me
to Yale, intuiting that if he and I and Bong Lian got together, something
synergistic would take place.  Meanwhile, thanks to e-mail, I had received
notes of a talk Ezra Getzler had given in Sydney which in turn led me to
learn more of his work in progress with J.D.S. Jones and that of
Ginzburg-Kapranov.
When I told Peter May of this renaissance in operad theory in
these several applications, he responded with a preprint \cite{KM}
where the general theory is advanced, precisely by the consideration of
partial algebras over operads, analogous to the partial operads
 which are essential to Huang's point of view.

Today I  will try to do what physicists call a `review', a survey talk
trying to bring all these strands together.

\{{\it Post conference:}
This talk was a preliminary effort,
based on rough drafts which had arrived only in the previous week.
Things  have not let up since I gave the talk;
I will do my best to make this report up to date as of the day submitted.
For historical purposes, remarks unknown to me in November 1992 are
inserted as is this remark.  A major influence during the development
of the talk and this paper has been provided by seminars at the
University of Pennsylvania and especially by my co-conspirators:
Y.-Z. Huang, T. Kimura and A. Voronov. \}

Let me begin, however, with a topic where some time ago I was able to
give a mathematical interpretation of the corresponding  physics.
 \vskip3ex
\Section{1.  Closed string field theory}
\vskip1ex
When particles are conceived of as points, much of particle physics can be
described in terms of Lie algebras and their representations.
Closed string field theory, on the other hand,
leads to a generalization of Lie algebra which arose naturally within
mathematics in the study of deformations of algebraic structures \cite{SS}.
It also appeared in work on higher spin particles \cite{BBvD}.
Representation theoretic analogs arose in the mathematical analysis of
the Batalin-Fradkin-Vilkovisky approach to constrained Hamiltonians \cite{S6}.
\par
         String field theory is multi-layered, often presented as involving
topology, geometry, algebra and analysis, especially analysis in the
sense of Riemann surfaces.  The bottom layer is the topology of string
configurations.

The obvious picture of a closed string is that of a
closed curve in a (Riemannian) manifold $M$.
The first subtlety
one encounters with this picture is that the physics is often
described in terms of a parameterization of such a curve, e.g. a
map of the standard circle $S^1$ (parameterized from $0$ to $2\pi$) into $M$,
but the physics should not
depend on the parameterization, although expressed in terms of it.
Think of arc length of a parameterized curve.  Thus the space of closed strings
$\cal C$ can be described as the space of equivalence classes
(under reparameterization) of maps of the standard (parameterized) circle
into $M$:
$$
Map(S^1, M)/Diff^+S^1.
$$
 We begin by reviewing the joining of two strings to form a third.
The picture is the familiar `pair of pants':
\vskip25ex
\centerline{Figure 1}
\pp
The details in terms of parameterization extend a method
  due to Lashof \cite L in the case of based loops and to Witten \cite {W1}
 for strings.  The idea is that two closed strings Y and Z join to form
a third $Y*Z$ if a semi-circle of one agrees with a reverse oriented
semi-circle of the other. (Notice this avoids Witten's marking of the
 circle.)
 The join $Y*Z$ is formed from the complementary
semi-circles of each.  To be more precise,
 consider the configuration of three arcs $A_i, i= 0,1,2$ with the three
 initial points identified and the three terminal points identified, as
 in a circle together with a diameter.
\vskip25ex
\centerline{Figure 2}
\pp

 (One is tempted to call
this a (theta) $\Theta$ curve, but string field theory is likely to involve
 theta functions in the sense of number theory; there's enough confusion of
 terminology already!) To emphasize the symmetry and to fix parameterizations,
 consider the arcs to be great semi-circles on the unit 2-sphere in ${\Bbb
R}^3$
 parameterized by arc length from north pole to south pole.
 Denote the union of the three arcs by $\Theta$. Denote by $\bar A_i$
 the arc parameterized in the reverse direction. Let $C_i$ denote
any isometry $C_i:S^i \righthookarrow \Theta$ which agrees with $A_j$
on one semi-circle and with $\bar A_k$ on the other for a cyclic
permutation $(i,j,k)$ of $(0,1,2)$.  (Up to rotation, $C_i$ is
$A_j$ followed by $\bar A_k$.)

Given any map $X: \Theta \rightarrow    M$,
let $X_i = X\compose C_i$ and think of $X_0$ as the fusion of $X_1$
and $X_2$.
\null
\vskip25ex
\centerline{Figure 3}
\pp
 When physicists speak of closed string {\it field} theory, I would
like to think
they are referring to `fields' which are some sort of functions, or more
generally some sort of sections of some bundle (unspecified) on
the space $\Cal C$ of closed strings, $Map(S^1, M)$.
Actually  Zwiebach (cf.
\cite {SZ}, \cite {KKS}, \cite K, \cite{Wies},  \cite{WZ}, \cite Z)
stipulates that the string fields
$\phi_1,\phi_2, \cdots$ are elements of
${\Cal H},$ a Hilbert space of a combined conformal
field theory of matter and `ghosts'.  The presence of ghosts indicates
the presence of a `BRST operator', which corresponds to the exterior
derivative along the leaves of the reparameterization orbits.

\vskip1ex
A convolution product for
 $\phi,\psi$ function(al)s on $\cal C$, the space of closed strings, is defined
in terms of such maps (although the details are unimportant
for this paper):
\vskip1ex
 Define the {\bf convolution product} $\phi * \psi$ as follows:

 $$
 (\phi*\psi)(X_0) = \int \phi(X_1)\psi(X_2) dX
 $$
\noindent
where the integrals are  over all isometries $C_i$ and over
all maps $X:\Theta \rightarrow M$ such that $X_0 = X\compose C_0$.
Thus $\phi*\psi$ depends on all ways
of {\bf decomposing} $X_0$ into two loops $X_1$ and $X_2$. (The
range $M$ is `space' and not `space-time' so there is no problem with
a Lorentz metric.
  In the case of the standard metric on $M = R^d$, notice
that this integral is over paths from $X_0(\theta)$ to $X_0(\theta + \pi)$
and hence is well defined as a Brownian bridge.)
\vskip1ex
Given a product, we ask immediately for its algebraic properties.
With an appropriate grading, the convolution product is graded
commutative but is nothing like associative; rather it comes close
to satisfying a graded Jacobi identity.  For this reason, we change notation
and
define
$$
\lbrack\phi, \psi\rbrack = \phi*\psi.
$$
For three fields $\phi_i, i = 1,2,3,$ there exists a trilinear
$\lbrack \phi_1, \phi_2, \phi_3 \rbrack$ such that
$$
\align
(1)\ \ \ [\p_1, [\p_2, \p_3]] \pm [[\p_1, \p_2], \p_3]&\pm [\p_2, [\p_1,
\p_3]]\\
&= d_1[\p_1, \p_2, \p_3] \pm [d_1\p_1, \p_2, \p_3]\pm[\p_1, d_1\p_2, \p_3]
\pm [\p_1, \p_2, d_1\p_3].
\endalign
$$
where $d$ is the vertical differential along the reparameterization orbits.
The form of the (super) Jacobi identity used here
is equivalent to other standard forms (e.g. the cyclic form) since
the bracket is super anti-commutative.  In more general algebras,
it will not be equivalent to other standard forms; this derivation
form has the easiest set of signs to remember and is the appropriate
one when considering operators.
\par
{}From this equation,
 we see that the Jacobi identity holds not strictly but rather modulo
the right hand side.  {\it In physical language, the Jacobi identity holds
modulo a BRST exact term.} In the language of homological algebra,
$d_3$ is a chain homotopy,  so we say that $(V,[\ ,\ ])$ satisfies the Jacobi
identity {\it up to homotopy} or $(V,d_1,[\ ,\ ],[\ ,\ ,\ ])$ is a
{\bf homotopy Lie algebra}.
\par
The genesis of this trilinear is an interpretation which sees the theta
curve as imbedded in a world sheet, which led several physicists,
 starting with Kaku \cite K, to consider a tetrahedral configuration in which
 the perimeter of each face is regarded as isometrically a circle (to be
mapped  via a closed string).
This requires that that pairs of opposite edges have equal lengths, say
$(a_1, a_2, a_3)$,
 with $0 \leq a_i \leq \pi$ and $\Sigma a_i = 2\pi$.  Denote such
a tetrahedron by
$\Delta^3(a_1, a_2, a_3)$.
Now consider an orientation preserving
$C_i$ taking $S^1$ isometrically to the boundary
of the $i$-th face of $\Delta^3(a_1, a_2, a_3)$.
\vskip30ex
\centerline{Figure 4}
\pp
My interpretation of the trilinear is then:
$$
[\p_1,\p_2,\p_3](X_0) =
1/6 \ \Sigma\int\int \p_{i_1}(X_1)\p_{i_2}(X_2)\p_{i_3}(X_3)dX
$$
where the sum is over all permutations $(i_1,i_2,i_3)$ and the integration
with respect to $dX$ is over all maps
$X:\Delta^3(a_1, a_2, a_3) \rightarrow M$ such that $X\circ C_0 = X_0$,
for all $\Delta^3(a_1, a_2, a_3)$
and  then the integration is over all isometries $C_i, i = 1,2,3$
and the space of all $\Delta^3(a_1, a_2, a_3)$.
The integral with respect to $dX$ is less standard than a Brownian
bridge, but has been addressed carefully
by Wiesbrock \cite{Wies}.  The homotopy algebra we are concerned with is
carried
by the more ordinary integration over the moduli space with isometries.
That the trilinear does indeed satisfy (1) follows
from an application of Stokes' Theorem, the right hand side arising as the
boundary terms when one of the $a_i$ equals $\pi$ (cf. Figure 5 ).
\vskip3ex
\Section{2.  Restricted Polyhedra}
 \vskip1ex
To proceed further, it proved necessary to consider polyhedra of
more than four faces.
 Extension to polyhedra with 5 faces was worked out by Saadi and
 Zwiebach \cite {SZ} and their lead was carried through to general polyhedra
 by Kugo, Kunitomo and Suehiro \cite {KKS}.  Here  `polyhedron' refers to a
cell
 decomposition of the oriented 2-sphere in which each face (=2-cell) has
boundary (perimeter) consisting of
 a finite number of edges (1-cells).  Each face and hence its perimeter
carries the orientation induced from $S^2$.
   The polyhedra are restricted
 geometrically in that each edge is assigned a length such that:
\par  \quad\quad (2.a) Saadi and Zwiebach: the perimeter of each face has
length $2 \pi$ (which implies each edge has length $\leq \pi$), and
\par   \quad\quad (2.b)  Kugo, Kunitomo and Suehiro: any simple closed edge
path has length $\geq 2\pi$.
\vskip2ex
(At the conference, Bonahon called my attention to the remarkable
fact that {\it exactly} these restrictions are the hypotheses for
Rivin's Theorem which has just recently appeared in Bull AMS, though
based on his earlier work \cite R.)
\vskip2ex
{\bf
Theorem:}
(Characterization of ideal convex polyhedra)
The dual polyhedron $P^*$ of a convex ideal
polyhedron $P$ in $H^3$ satisfies the following conditions (in terms of
`weights' $w(e^*)$ for edges of $P^*$:
\par  {\bf \quad\quad\quad Condition 1.} $0<w(e^*)<\pi$ for all edges $e^*$
of $P^*$.
\vskip2ex
{\bf \quad\quad\quad Condition 2.} If the edges
$e_1^*, e_2^*, \dots, e_k^*$
form the boundary of  a face of $P^*$, then $$w(e_1^*)+w(e_2^*)+\dots +
w(e_k^*) = 2\pi.$$
\par {\bf \quad\quad\quad Condition 3.} If $e_1^*, e_2^*, \dots, e_k^*$
form a simple circuit which does not bound a face of $P^*$, then
$$w(e_1^*)+w(e_2^*)+\dots + w(e_k^*) >  2\pi.$$
Conversely, any abstract polyhedron
$P^*$ with weighted edges satisfying the conditions 1-3
is the Poincar\'e dual of a convex ideal polyhedron $P$ with
the exterior dihedral angles equal to the weights.
\vskip2ex
We return to restricted polyhedra and consider the ``moduli'' spaces thereof in
some detail. There is only one restricted trihedron, $\Theta$.
\vskip1ex
For tetrahedra, the restrictions are precisely
that opposite edges have equal lengths, say $(a_1, a_2, a_3)$, with $0
\leq a_i \leq \pi$ and $\Sigma a_i = 2\pi$ (Figure 4).  In other words, the
``moduli'' space of restricted tetrahedra is given by the union of two
2-simplices (one for each orientation of the tetrahedron) with vertices
in common.
\vskip25ex
\centerline{Figure 5}
\pp
More generally, let ${\cal V}_N$ be the ``moduli'' space of all
restricted $(N+1)$-hedra $P$
with an arbitrary ordering of the faces from 0 to N.
(We have tried to keep our notation close to Zwiebach's; in particular,
his $n = N + 1$.  As a space, ${\cal V}_N$ is given the topology of the local
coordinates which are the edge lengths - in fact, this is a cell
decomposition.  This moduli
space ${\cal V}_N$ of all restricted $(N+1)$-hedra as defined is
manifestly a union of cells indexed by the topological type of the
polyhedron (and the ordering).  For a given topological type,
the restrictions 1)
and 2) with edge lengths $ > 0$ describe an open convex subspace (polytope)
${\cal V}_N$ of $R^E$ where $E$ is the number of edges of the
$(N+1)$-hedron.  Strictly speaking, when an edge length goes to zero,
the topological type of the $(N+1)$-hedron changes and thus two cells
are glued together along a cell of one lower dimension.
\vskip30ex
\centerline{Figure 6}
\pp
 Thus ${\cal V}_N$ is
described as a finite cell complex in which each cell has boundary
composed of a finite number of cells of lower dimension.  The cells of
maximal dimension correspond to 3-valent polyhedra and the dimension of
these cells is $2N-4 = 2(N+1) - 6$, with faces corresponding to $(N+1)$-hedra
with one 4-valent vertex, etc. when an edge length goes to zero.
\pp
I have emphasized the topology of $\Cal V_N$ as a cell complex.  Zwiebach
instead regards $\Cal V_N$ as a subspace of the traditional moduli space
$\Cal M_{0,N+1}$
of Riemann spheres with $N+1$ punctures by filling in each
face of the polyhedron with a punctured disk realized in terms of a metric of
minimal area \cite Z.  Thus $\Cal V_N$ can be regarded as a subspace of
$\Cal M_{0,N+1}$
``cut-off'' from the degenerations of Riemann spheres having double points
appearing in various compactifications of
$\Cal M_{0,N+1}$ \cite {Kn} \cite D \cite {FM}.

\pp
{\bf Decorations}
\pp
We are going to ``decorate'' such polyhedra with
specific  parameterizations $C_i:S^1
\hookrightarrow P$ which are isometries with the perimeter of the i-th face
- analogous to a local coordinate at one of $(N+1)$ punctures on a
Riemann surface of genus 0.
\vskip1ex
Let ${\cal P}_N$ denote the space of ordered restricted
$(N+1)$-hedra with such specified isometries $C_i:S^1 \hookrightarrow
P$.  It is naturally a bundle over ${\cal V}_N$ with fibres $(S^1)^{N+1}$
(homeomorphic to ${\cal V}_N \times (S^1)^{N+1}$).
We will use the notation $\bar P = (P,C_0, \dots ,C_N) \in {\cal P}_N.$
Note that Zwiebach's $\Cal P_{0,n}$ is more fully decorated with arbitrary
complex local coordinates at the punctures.  Though natural in conformal field
theory, these more general coordinates are not needed for the structures we
consider.

\vskip3ex
\Section {3. The N-ary operations}
 \vskip1ex
  We are now ready to define
N-ary operations, multi-linear `brackets', which are key to the closed
string field theory of KKS and Zwiebach, in that they satisfy the
crucial relation
$$
(J_N)\hskip8ex \16N
$$
where the signs are the usual one for interchanging graded objects.

\def\H{\cal H}

\pc Here is a revisionist version of my description of  a portion of closed
string
field theory, which hopefully has benefitted from discussions with Alexander
Voronov (since my talk).
The data consists of a Hilbert space $\cal H$, graded by `ghost number',
an integer; in fact, $\cal H$ is a differential graded Hilbert space;
the differential is of degree $+1$, is denoted $Q$  and is called a
BRST operator.   For our purposes,
the essential aspect of a conformal field theory is that it
provides a chain map
from $\H^{\otimes N+1}$ to the ordinary
differential forms on $\cal P_N$:
$$
\H^{\otimes N+1} \to \Omega^{\bullet}( \cal P_N).
$$
(As so often in physics, $\H$ is identified
with its linear dual $\H^*$ and similarly for iterated tensor products.)\}
\par
 If the result is then pulled down to $|Cal V_N$ via a section and
integrated over $\cal V_N$, we obtain
$$
<\psi_0\dots\psi_N> \text{\quad for\quad} \psi_i\in \H
$$
which, if the ghost numbers of the $\psi_i$ have the correct
total, is a number, called the {\bf correlator} or {\bf $N+1$-point function}
 of the $\psi_i$.
The  $N$-fold brackets are then determined by the correlators or vice versa
via:
$$
<\psi_0\dots\psi_N> = <\psi_0 \vert [\psi_1,\dots,\psi_N]>
$$
or
$$
[\psi_1,\dots,\psi_N]:=\Sigma \psi^{\alpha}<\psi_{\alpha}\psi_1\dots\psi_N>.
$$
\vskip3ex
\Section {4. The bracket relations}
\vskip1ex
The cell complex ${\cal P}_N$ does have a ``boundary'' corresponding to
saturation of the inequalities (2.b).  KKS refer to such polyhedra as {\bf
critical}, i.e. if there is an edge path
enclosing at least two faces and of length precisely $2\pi $.
 Separating the polyhedron $P$ along this edge path produces two
restricted polyhedra $Q$ and $R$.
\vskip35ex\par\noindent
\centerline{Figure 7}
\pp

The separation can be regarded as giving a partition of the set $\{0,\dots
,N\}$
or, if the faces of $P$ are ordered,  as an {\bf unshuffle} giving induced
orderings on $Q$ and $R$.
(An {\bf  unshuffle} of $\lbrace 1,\dots ,n\rbrace$
is a permutation that keep $i_1,\dots ,i_j$ and $i_{j+1},\dots ,i_n$
in the same relative order.  A {\bf shuffle} of two ordered sets (decks
of cards) is a permutation of the ordered union which preseves the order
of each of the given subsets; an {\bf unshuffle} reverses the process, cf.
Maxwell's demon.)
\vskip2ex
\vskip1ex
Conversely, if we have two decorated restricted polyhedra $Q$ and $R$,
we can form a connected sum $Q \# R$ by deleting two faces
$F_Q \in Q$ and $F_R \in R$ and identifying
their perimeters by an orientation reversing isometry. (Since we regard
$S^1$ as parameterized once and for all from $0$ to $2\pi,$
the orientation reversing isometry is uniquely specified by requiring
that the images of $0$ coincide.)
Generically, the result will be a restricted polyhedron $P$ (although
occasionally the identification may produce vertices of valence greater
than 3).  If the faces of $Q$ and $R$ are ordered, we can define such a
connected sum for each shuffle.
\vskip1ex
Thus we can describe boundary ``facets'' of ${\cal P}_N$ by inclusions
$$
 {\cal P}_Q \times_{S^1} {\cal P}_R  \hookrightarrow {\cal P}_N
 $$
 where $Q + R = N + 1$, the inclusions being indexed by $(Q,R)$-shuffles
and the quotient by $S61$ corresponding to the fact that one decorated
polyhedron $P$ can result from different decorations before identification
as long the orientation reversing isometry is the same.
\pp
Similarly, we can describe boundary ``facets'' of ${\cal V}_N$ by inclusions
$$
 {\cal V}_Q \times {\cal V}_R\times S^1 \hookrightarrow {\cal V}_N
 $$
 where $Q + R = N + 1$, the inclusions being indexed by $(Q,R)$-shuffles
and the factor of $S^1$ specifying the possible rotations possible in
identifying isometrically the unparameterized perimeters.
\par
The proof of the defining relation $(J_N)$ is again essentially an
application of Stokes' Theorem.  This time
the  boundary term can be written as
$$
\integral_{{\cal V}_R}\integral_{{\cal V}_Q}
$$
giving rise to the terms
$$
[[\phi_{i_1},\dots,\phi_{i_Q}],\phi_{i_{Q+1}},\dots,\phi_{i_N}].
$$

  \vskip3ex
\Section{5.  Strongly homotopy Lie algebras}
\vskip1ex
\def\bw{\bigwedge}
\def\16N{d_1\lbrack\phi_1\dots\phi_N\rbrack \pm \Sigma_1^N\pm\lbrack\phi_1
\dots
d_1 \phi_i\dots\phi_N\rbrack = \Sigma^{N-1}_{J=2}\pm\lbrack\lbrack\phi_{i_1}
\cdots \phi_{i_J} \rbrack\phi_{i_{J+1}}\cdots\phi_{i_N}\rbrack}
%\def\cite#1{[{\bf #1}]}
% \def\myspace{\hskip 1em\relax}
%\def\cite#1{[{\bf #1}]} \def\myspace{\hskip 1em\relax}
%\pageno=1
% \headline={\ifnum\pageno>1 \tenbf
% SHLie algebra for physics \hss Lada-Stasheff\else\hfil\fi}
\par
To make sense out of the formula $(J_N)$, we re-examine the concept of
Lie algebra,
which can be expressed in several different ways. The
most familiar are: in terms of generators and relations and in terms of a
bilinear ``bracket'' on a vector space $V$ satisfying the Jacobi identity.
A much more subtle
description appears in the homological study of Lie algebras
and is implicit in the somewhat  more familiar dual
formulation of the Chevalley-Eilenberg
cochain complex for Lie algebra cohomology \cite{CE}.
  We can deal directly with
the vectors rather than with the multi-linear `forms' at the expense
of introducing a new point of view and consideration of skew-symmetric
tensors ({\it poly-vectors}).
\par
\vskip2ex\noindent
A Lie algebra is equivalent to the following data:
\vskip0.5ex
A vector space V (assumed finite dimensional for simplicity  of exposition).
\vskip0.5ex
The skew (or alternating) tensor products of V, denoted
$$\bw     V = \{\bw ^nV\},$$
with $\bw^0 V = k,$ the field of scalars, typically the reals, $\Bbb R$ or
the complex numbers, $\Bbb C$.
\vskip0.5ex
A linear map
$$d:\bw     V \rightarrow \bw     V$$
 which lowers $n$ by one and is a {\it co-}derivation
determined by $d\vert\bw^2 V$
such that $ d^2 = 0$.
\vskip0.5ex\noindent
(That $d$ is a {\bf co-derivation}
determined by $d\vert\bw^2 V$
means just that $d(v_1) = 0$ and
$$d(v_1\wedge \dots \wedge v_n) = \underset{i<j}\to\Sigma (-1)^{i+j}
d(v_i\wedge v_j)\wedge v_1\wedge \dots \hat v_i \dots \hat v_j \dots
\wedge v_n
$$
where $\hat v_i$ denotes the deletion of $v_i$.)

\par
It may not be immediately obvious, but $d$ restricted to $\bw ^2$ is
to be interpreted
as a bracket: $d(v_1\wedge v_2) = [v_1, v_2]$ and $d^2 = 0$ is equivalent to
the
Jacobi identity.
\par
The generalization we  need `up to higher homotopy of all orders' is
comparatively straightforward
from this point of view of the skew tensor powers of $V$;
an {\bf sh Lie algebra (strongly homotopy Lie algebra)} \cite{LS}
is equivalent to a  straightforward generalization in which d is
 replaced by  a coderivation
$$D = d_1 + d_2 +d_3 + \dots$$ where $d_i$ lowers $n$ by $i-1,$ in particular,
$d_n(v_1\wedge \dots \wedge v_n)\in V$.  The only subtlety is in the signs.
The vector space $V$ is a graded vector space and the `skew symmetry'
is better expressed as graded symmetry where the grading has been
shifted by $1$.  We denote this by
$$
\bw sV.
$$
\par
(A further generalization in which there is a term $d_0$ interpreted as
a fixed ``background'' field occurs in Zwiebach's investigation of background
dependence, but for this there is no mathematical precursor, to our knowledge.)
\par
We say that $D$ is a coderivation to summarize the several conditions:
$$
d_j(v_1\wedge \dots \wedge v_n) = \Sigma \pm d_j(v_{i_1}\wedge\dots\wedge
v_{i _j})\wedge v_{i_{j+1}}\wedge \dots\wedge v_{i_n}
$$
where the sum is over
all  unshuffles of $\lbrace 1,\dots ,n\rbrace$.
\vskip2ex
Notice that the old $d$ corresponds to $d_2$ since
$d(v_1\wedge v_2) = [v_1, v_2]\in V.$
On the other hand, for the new D, the component $d_2$
no longer is of square zero by itself and hence corresponds to a bracket which
does  NOT necessarily satisfy the Jacobi identity.  Let us look in detail at
what can happen instead:
\vskip1ex
Expand $D^2 = 0$ in its homogeneous components and set them separately
equal to zero.  We have then:
$$
0)\quad  d_1^2 = 0
$$
so $(V,d_1)$ is a complex or differential (graded) module.  (Typically
$d_1$ raises (or lowers) degree by 1.)
$$
1) \quad d_1 d_2 + d_2 d_1 = 0
$$
so, with appropriate sign conventions,  $d_2$
gives a bracket $[v_1, v_2]\in V$ for which
$d_1$ is a derivation.
$$
2) \quad d_1 d_3 + d_2 d_2 + d_3 d_1 = 0
$$
or equivalently
$$
d_2 d_2 = -(d_1 d_3 + d_3 d_1).
$$
If we further adopt the notation:
$$
d_3(v_1\wedge v_2\wedge v_3) = \lbrack v_1,v_2,v_3\rbrack,
$$
then we have
$$
[[v_1, v_2], v_3] \pm [[v_1, v_3], v_2] \pm [[v_2, v_3], v_2] =
- d_1[v_1, v_2, v_3] \pm [d_1v_1, v_2, v_3]\pm[v_1, d_1v_2, v_3]\pm [v_1, v_2,
d_1v_3].
$$
In the language of homological algebra,
we say that $(V,d_1,d_2,d_3)$ is a
{\bf homotopy Lie algebra}.  The adverb ``{\bf strongly}'' is added to refer
to the other $d_i$, the higher homotopies.
\par
If we adopt the notation that $d_1 = Q$ and in general:
$$
d_n(v_1\wedge \dots\wedge v_n) = \lbrack v_1,v_2,\dots ,v_n\rbrack,
$$
then the appropriate homogeneous piece of $D^2 = 0$ is (up to sign conventions
and up to some constants
related to conventions on the definition of $\bw V$) precisely the equation
$(J_N).$
\par
In the higher spin particle algebra of \cite{BBvD}, variations $\delta_
{\epsilon}$ do not respect a strict bracket $[\epsilon_1, \epsilon_2]$
but rather an sh Lie structure on the space of $\epsilon$'s.  In the
Batalin-Fradkin-Vilkovisky operator for constraints forming an `open' algebra
with structure functions, one sees a similar structure \cite {S6}.
\par
\pc \ Schechtman has informed me that Drinfel'd described sh Lie algebras in
essentially
these terms in a letter to Schechtman in 1988.\}
\vskip3ex

\def\R{{\Bbb R}}
\def\C{{\Bbb C}}
\def\OP{\Cal O}
\def\SS{\Sigma}
\def\Ainf{$A_{\infty}$}
\def\newcommand{\def}

\Section{6.  The operad structure}
\pp
We return to the geometry underlying this algebra in closed string field
theory. From now on, the face labelled 0 of a restricted polyhedron
will play a distinguished role.
Think of inputs going in through the other n faces with face 0
reserved for output.
Then decorated restricted polyhedra (elements
of ${\cal P}$) can be combined exactly as rooted trees can by
``grafting'' roots to branch tips.
The sequence of spaces $\Cal P_k$ provides an example of May's notion of an
operad \cite{M1}: this is a sequence of spaces $\OP(k)$ satisfying
certain conditions modelled on those satisfied by $Map(X^k,X)$
with the obvious action of the symmetric group $\SS_k$.

{\bf A space $X$ is acted on by an operad} $\OP$ means
there is a sequence of $\SS_k$-equivariant maps
$\theta_k:\OP(k)\times X^k@>>>X$ satisfying certain compatibility
conditions. More generally, operads can be defined in any symmetric
monoidal category (so that $X^k$ makes sense)\cite{Ke},
for example, the category of topological spaces, the category of graded vector
spaces (with tensor product), or the category of chain complexes (once
more with tensor product).
\vskip1ex
Definition. An {\bf operad} $\OP$ in such a category is a sequence
$\OP(k)$, $k\ge1$, of objects with
$\SS_k$- action
and maps
$$
\gamma:
\OP(k) \times \OP(j_1)\times\dots\times\OP(j_k) @>>> \OP(j_1+\dots+j_k) ,
$$
satisfying conditions of `operad-associativity'
for iterating $\gamma$ and ``operad-symmetry'' (respecting the
$\SS_k$ action)  - which are obvious in the following example built from
$Map(X^k,X)$.
\vskip1ex
THE basic example, in any such category, is the {\it endomorphism operad}
$$
\Cal E_X = \lbrace \Cal E_X(k) = Map(X^k, X)\rbrace
$$
with $\gamma$ given by composing a map with $k$ ordered inputs with the
outputs of $k$ (other)
maps in the usual way.
\vskip1ex

Two very important  operads related to ordinary algebra are
$$
\Sigma =\lbrace \Sigma(k) = \SS_k\rbrace
$$
 with the structure map $\gamma$ given by a generalization of
wreathe product, and
$$
\Cal N = \{k\}
$$
interpreted as singletons.
\par
Notice that the distinct ways of multiplying elements
$(x_1,\dots,x_k)$ of an associative algebra can be indexed by
the elements $\pi\in\SS_k$, namely the iterated products
$x_{\pi(1)}\dots x_{\pi(k)}$.
Thus an action of $\Sigma$ on a set $X$ makes $X$ an associative
monoid. Similarly, an action of $\Cal N$ makes $X$
a commutative associative monoid since all of the
$x_{\pi(1)}\dots x_{\pi(k)}$
are to be equal.
\par
Operads were originally invented \cite{M1} for the study of iterated
(based) loop spaces: for two
excellent overviews of this theory, see Adams \cite{Ad} and May
\cite{M2}.
Before that invention (and hence without the name), I created  an operad
\cite{S4, S3} that made explicit the
higher homotopies required of the multiplication on an H-space for it to be
homotopy equivalent to a loop space.  I introduced a sequence of convex
polyhedra $K_k$ (which have come to be known as {\bf associahedra}), of
dimension $k-2$, with the property that a connected space $X$ has the
homotopy type of a loop space if and only if there is a sequence of maps
$$
\theta_k : K_k \times X^k @>>> X
$$
satisfying certain compatibility conditions. Such a space is called an
\Ainf-space. The associahedron $K_2$ is a point, so that $\theta_2$ gives
$X$ the structure of an H-space.  Furthermore, this product is homotopy
associative, in the sense that $K_3$ is an interval, and the two products
$(ab)c$ and $a(bc)$ correspond to the two endpoints of $K_2$, thus
$\theta_3$ gives a canonical homotopy between $(ab)c$ and $a(bc)$.
The associahedron $K_4$ is the now familiar pentagon; $K_5$ is
pictured in Figure 8 in a visualization I owe to John Harer.
\vskip35ex
\centerline{Figure 8}
\pp
The decorated restricted polyhedra (elements
of ${\cal P}$) form an operad in the obvious way; just as the
isometries $C_i$ were used to `glue' two restricted polyhedra
to form a third, so the isometries allow us to glue $k$ polyhedra
in $\cal P_{j_i}$ to a $k+1$-polyhedron in $\cal P_k$ to form a
$(j_1+\dots+j_k +1)$-polyhedron.
\par
For n-fold iterated loop spaces, Boardman and Vogt \cite{BoVo}
introduced a useful sequence of operads,
the little $n$-cubes operads $\CC_n(k)$, which have the homotopy type
of the configuration spaces $ F({\Bbb R}^n,k)$
of $k$-tuples of distinct points in
$\R^n$; a space with an action of the operad $\CC_n$ is called an
$E_n$-space and has the homotopy type of an $n$-th loop space, at
least if it is connected. Certain cases are of particular interest:
$n=1$ recovers the theory of \Ainf-spaces, $n=\infty$ leads to infinite
loop spaces and $n=2$ is intimately related to the braid groups
$\BB_k$, since $\config /\SS_k$ has the homotopy type of
$K(\BB_k,1)$.
\par
Henceforth we will be concerned only with $n = 2$ for which we have
the following notation and formal definition:
\par
Definition: {\bf The little squares operad} $\cal C_2$:
Let $\CC_2(k)$ be the space of all maps
from $\coprod_{i=1}^kI^2$ to $I^2$ which are
affine on each coordinate of each square
and such that the images of the $k$ squares are
disjoint. The symmetric group $\SS_k$ acts by permuting the cubes in
the domain of the map. The operad structure is given by the maps
$$
\gamma: \CC_2(k) \times \CC_2(j_1) \times \dots \times
\CC_2(j_k) @>>> \CC_2(j = j_1+\dots+j_k)
$$
defined by
$$
\gamma (c,d_1,\dots,d_k)(x_1,\dots,x_j)
= c(d_1(x_1,\dots,x_{j_1}),\dots,d_k(x_{j_1+\dots+j_{k-1}+1},\dots,x_j)) .
$$

Closely related is the {\bf geometric little disks operad}
 $\cal D = \{\cal D(k)\}$\cite{M1}:
Let $D$ be the unit disc in $\C$.  Let $\Cal D(k)$ be the space of all
maps $(z_1,\dots,z_k)$ from $
%\displaystyle
\coprod_{i=1}^kD$ to $D$ which
are obtained by dilation and translation   such that the
images of the $k$ discs are disjoint. The operad structure is again
defined by composition.

We can think of the little disks as being cut out of the standard disk
and hence think of certain Riemann surfaces of genus $0$ with $k+1$ {\it
parameterized} boundary components.
\par
The decorated restricted polyhedral operad $\cal P$
is very similar except that, as Albert Schwarz remarked, they are
`Riemann surfaces with boundary only' and the boundary components have
various (isometric) parameterizations.  We can similarly `decorate'
the elements of $\cal D$ by allowing rotations in the maps
$\coprod_{i=1}^kD$ to $D$; we denote the resulting operad by $\tilde\cal D.$
\par
Huang \cite{Hu} adopts a related point of view, that of Riemann surfaces
with punctures and local coordinates, more appropriate to {\it sewing}
rather than {\it glueing}, with the crucial difference that he constructs
a {\it partial} operad, $\cal H$, i.e. the structure map $\gamma$ is not
always defined but when it is, the compatibility does hold. (Further details
are in Section  9.)
\par
The undecorated operads $\CC$ and $\cal D$ have the homotopy type
of the non-operad of configuration spaces
$ F({\Bbb R}^2,k)$.  The decorated operads $\Cal P$ and $\tilde
\cal D$ have the homotopy type of torus bundles over
the non-operad of configuration spaces
$ F({\Bbb R}^2,k)$.
\vskip3ex
\Section {7. Operads via homology or chains}
\pp
Operads in one category give rise to operads in another by applying
a suitable functor.  In particular, for an operad in a category of
topological spaces, homology with field coefficients is such a functor.
(Getzler \cite{Get} starts with smooth or complex spaces and Huang \cite{Hu}
with complex manifolds.)  Until quite recently, an operad to characterize
Lie algebras had not been given in as formal a fashion as for
associative  or commutative associative
algebras, but, as pointed out by Getzler and Jones \cite {GJ}, there
was one implicit in the work of Fred Cohen on the
homology of configuration space.  Arnol'd
\cite{Ar} and later but independently Fred Cohen \cite{C1}
determined the homology of $\config$.  Cohen \cite {C2} also provided the
following related result of direct relevance for us.
For any graded vector space $V$, let $sV$
denote the graded vector space such that $(sV)_{n+1} = V_n$;
the operator $s$ shifts the grading by $1$.
\pp
{\bf
Theorem:} For homology with coefficients in a field and any graded vector space
$V$ over that field,
$$
s\bigoplus_k H_{k-1}(\config) \otimes_{\Sigma_k} V^{\otimes k}
$$
is isomorphic to the free graded Lie algebra generated by $sV$.

Thus, in the category of graded vector spaces, we can regard
$$
\cal L = \{\cal L(k) = H_{k-1}(\config)\}
$$
as an operad $\cal L$ such that $\cal L$ acting  on a graded vector
space gives $sV$ the structure of a graded Lie algebra or,
alternatively, it is sometimes said that $V$ has a Lie bracket of degree $-1$
or in physspeak, an `anti-bracket'.
\par
 An alternate approach to sh Lie algebras (inspired by work of Beilinson and
Ginzburg \cite {BG1}) has been developed by Hinich and
Schechtman \cite {HS2}. Rather than use $\cal L$ as described above,
they describe it in terms of certain sub-modules of the free Lie algebras
on $n$ variables and call it the `trivial Lie operad' $\cal L$.  `Weak
versions', equivalent to sh Lie algebras, are described in terms of operads
quasi-isomorphic to $\cal L$ and a universal one is constructed.
This universal one bears a nice relation to appropriate compactifications of
$\Cal M_{0,N+1}$, as explained further in Section 13.
In particular, this relates to a good chain complex model for $\config $.

\par
Similarly, Cohen's results show the full homology
$$
\cal G =\{\cal G(k) = H_*(\config)\}
$$
also forms an operad, which characterizes a {\bf Gerstenhaber algebra}
(a.k.a. {\bf braid algebra} in the terminology of Getzler and Jones).
The formal definition is:
\pp
Definition.  A {\bf Gerstenhaber or braid algebra} is a graded vector
space $V$ with a product
$ V^p\times V^q\to  V^{p+q}$ and a bracket
 $\{\ ,\}: V^p\times V^q\to  V^{p+q-1}$ such that

(a) $uv=(-1)^{|u||v|}vu$

(b) $(uv)w=u(vw)$

(c) $\{u,v\}=-(-1)^{|su||sv|}\{v,u\}$

(d) $\{u,\{v,w\}\}= \{\{u,v\},w\} +(-1)^{|su||sv|}\{v,\{u,w\}\}$

(e) $\{u,vw\}=\{u,v\}w +(-1)^{|su||sv|}v\{u,w\}.$
\pp
\pp
%\begin{example}
The homology of a $2$-fold loop space $H_\bull(\Om^2X)$ is a Gerstenhaber
algebra, although Cohen \cite{C1} does not
use that name; the product is the
Pontryagin product coming from the H-space structure of $\Om^2X$,
while the Lie bracket is called the Browder operation.  In fact,
if $X$ is the 2-fold suspension of a space $Y$, then
$H_\bull(\Om^2X)$ is the free Gerstenhaber algebra generated by the
homology of $Y$.
%\end{example}
\par
Gerstenhaber's creation of the first example of this structure
occurred in a purely algebraic context.
The Hochschild cohomology $H^\bull(A,A)$ of an associative algebra is a
Gerstenhaber or braid algebra; the product is the cup product,
while the bracket is
Gerstenhaber's \cite{Ger}. In the special case that $A$ is the
algebra of differentiable functions $\Cinf(M)$ on a manifold $M$, the
Hochschild cohomology was shown by Hochschild-Kostant-Rosenberg \cite{HKR}
to be naturally isomorphic to the space of multivectors
$\Gamma(M,\bigwedge^\bull TM)$. With this identification, the cup product
may be identified with the wedge product on $\Gamma(M,\bigwedge^\bull TM)$,
while the Gerstenhaber bracket may be identified with the
Schouten-Nijenhuis-Richardson bracket.
%\end{example}
\par
But what of the decorated operads?  What structure are they characterizing?
Before answering that in the terms of Getzler and Jones, let me first
comment on Zuckerman's talk.
\vskip3ex
\Section {8. The G-algebra of BRST string theory}
\pp
In the first talk of this conference, Zuckerman reported on his recent work
with Lian \cite{LZ} in which they develop the structure of a Gerstenhaber
algebra on the BRST cohomology of a chiral algebra.  To be precise, he
stated:
\par
{\bf Theorem  LZ 1:} The BRST cohomology of a chiral algebra (a.k.a.
chiral cohomology)
$H^*$ admits the structure of a Gerstenhaber algebra and the action of $b_0$
is a derivation of the Gerstenhaber bracket.
\par
In more detail, they construct a product $u\cdot v$
and a bracket $\{ u,v\}$ such that the BRST operator $Q$ is a derivation of
both the product and bracket.  Moreover:
\par

{\bf Theorem LZ 2.2:}  On the chiral cohomology $H^*$, we have
$u\cdot v: H^p\times H^q\to  H^{p+q}$ and
 $\{,\}: H^p\times H^q\to  H^{p+q-1}$ such that

(a) $u\cdot v=(-1)^{|u||v|}v\cdot u$

(b) $(u\cdot v)\cdot w=u\cdot(v\cdot w)$

(c) $\{u,v\}=-(-1)^{|su||sv|}\{v,u\}$

(d) $(-1)^{|su||sw|}\{u,\{v,w\}\}
+(-1)^{|sw||sv|}\{w,\{u,v\}\}
+(-1)^{|sv||su|}\{v,\{w,u\}\}=0$

(e) $\{u,v\cdot w\}=\{u,v\}\cdot w +(-1)^{|su||v|}v\cdot\{u,w\}$

(f) $b_0\{u,v\}=\{b_0u,v\}+(-1)^{|su|}\{u,b_0v\}$.

\pp
On the BRST complex itself (`off-shell' in physspeak),
all the identities
of a Gerstenhaber algebra hold up to homotopy.   Of course, it
is the structure of these homotopies that excites my interest.  Many of
these identities up-to-homotopy are the same as they are in the Hochschild
complex, as given by  Gerstenhaber \cite {Ger}.  A striking difference is that
the Hochschild product of cochains is strictly associative, while that of
Lian and Zuckerman is only associative up to homotopy:
%\begin{eqnarray}\lb{2.10}
$$
(b')\  (u\cdot v)\cdot w - u\cdot (v\cdot w) = \hskip30ex
$$
$$
 Qn(u,v,w)+n(Qu,v,w)+(-1)^{|u|}n(u,Qv,w)
+(-1)^{|u|+|v|}n(u,v,Qw)
$$
where $n$ is a trilinear operation defined using $b_1$.
\par
The Lian-Zuckerman homotopy versions of (a) and (c)-(e) above are:

$$
(a')\  u\cdot v=(-1)^{|u||v|}v\cdot u = Qm(u,v)+ m(Qu,v)+(-1)^{|u|}m(u,Qv)
$$
where $m$ is a bilinear operation again defined using $b_1$,
$$
(c')\  \{u,v\}+(-1)^{|su||sv|}\{v,u\}
= (-1)^{|su|}(Qm'(u,v)-m'(Qu,v)-(-1)^{|u|}m'(u,Qv))
$$
where $m'$ is built using the homotopy $m$ and $b_0$

$$(d')\  \{u,\{v,w\}\} = \{\{u,v\},w\} +(-1)^{|sv||su|}\{v,\{u,w\}\}
$$
if the antighost field is defined with care,

$$
(e') \{u,v\cdot w\}=\{u,v\}\cdot w +(-1)^{|su||v|}v\cdot\{u,w\}
$$
but
$$
\align
\{u\cdot v,w\}-u\cdot\{v,w\}-&(-1)^{|sw||v|}\{u,w\}\cdot v =\\
 (-1)^{|u|+|v|-1}( Qn''(u,v,w)-n''(Qu,v,w) -&(-1)^{|u|}n''(u,Qv,w)
-(-1)^{|u|+|v|}n''(u,v,Qw))
\endalign
$$
where $n''$ is a trilinear operation defined using $m'$ and the product.
\par
Notice that the two versions of (e') are different, but exactly as they are
in the Hochschild complex, or earlier in Steenrod's relation between
$\smile$ and $\smile_1$.  By simply skew-symmetrizing the bracket, (c') can
be replaced by (c); the price to be paid is that (d') then holds only
up to homotopy.  This however is preferable if higher homotopies are to be
studied as in string field theory.
\par
\par
Finally, the bracket and product are related via $b_0$:
$$
(f') = (f): (-1)^{|u|}\{u,v\}= b_0(u\cdot v)-(b_0 u)\cdot v
-(-1)^{|u|}u\cdot(b_0v).
$$
A remark about this
identity: it  clearly measures the failure of $b_0$
to be a derivation of the dot product. The same idea appears in the
Batalin-Vilkovisky
``anti-bracket'' formalism\cite{BaVi}, but in a seemingly different context.
In \cite{W2}, Witten showed that the
Batalin-Vilkovisky master equation
can be formulated using a certain fundamental differential operator
$\Delta$ in field space, together with an anti-bracket which
measures the failure of $\Delta$ to be a derivation of an
operator product. The $b_0$ operator here plays the role
of $\Delta$.  This extra operator on a Gerstenhaber algebra is
the signature of what Penkava and Schwarz and Getzler and Jones have named a
Batalin-Vilkovisky algebra, which we discuss below.
\par
These homotopies suggest that there is an fact a strong homotopy Gerstenhaber
algebra structure present - in a sense yet to be defined, but see below.
First, here is an alternate approach due to Y.-Z. Huang.
\vskip3ex
\Section {9. Partial operads and partial algebras}
\pp
Huang \cite{Hu, HL} considers a partial operad $\cal K$ (that is, the structure
maps are defined only on suitable subsets of the usual range, cf. \cite{Ste}),
 but one with analytic structure, cf. \cite {Get}.
His partial operad is the moduli space of
Riemann spheres with $n+1$ ordered punctures and a local
coordinate vanishing at each puncture, further restricted in that the
$0$-th puncture is negatively oriented and the other punctures are
positively oriented.
\par
The operad structure map $\gamma$ is defined by sewing Riemann spheres
with punctures and local coordinates with the orientations specified
above, the sewing being of the $0$-th puncture  of one to one of the positively
orinted punctures of the other.
The sewing of two such spheres with punctures and
local coordinates is defined by cutting disks
at the specified punctures using the local
coordinates and then identifying the boundaries of the disks using the
map $z\rightarrow 1/z$ . The sewing is defined only when the local
coordinate neighborhoods can be extended such that the disks we want
to cut are inside the extended neighborhoods, and there are no
other punctures inside these disks. Therefore $\gamma$ is only
partially defined.
\par
Huang considers vertex operator algebras (which correspond to chiral
algebras without ghosts or anti-ghosts) and establishes that a VOA-structure
on a vector space V implies  a projective action of
$\Cal K$ on $V$, except that
a double dual of $V$ is involved.
Conversely such an action on $V$ gives $V$ a VOA-structure.
Since the introduction of ghosts and
anti-ghosts leads to a super chiral algebra with the additional subtlety
only of some signs,  Huang's techniques also give  a projective
$\Cal K$ action on  the underlying vector space $V$ of an
appropriate BRST complex.
\pp
In the preprint of Kriz and May \cite{KM} ,
the general theory is advanced, not by the consideration of partial
operads but rather by the consideration of partial actions of an
operad, i.e.
a partial algebra over an operad $\cal O$ has similarly coherent maps
$$
\cal O(k)\otimes A_k\to A
$$
 where $A_k\subset A^k.$
\pp
Kriz and May explain: ``The original motivation of this paper was to show
how to construct May
algebras from partial algebras'', May algebras
(the terminology is due to Hinich and Schechtman in earlier work)
being algebras over a special class of $``E_{\infty}-$operads''.
``
This work was inspired by letters from Deligne to Bloch and May.  Deligne
suggested
that Bloch's Chow complex, which is a partial algebra, should give rise to
a quasi-isomorphic (graded) May algebra...'',  with an eye to its
derived category of modules providing a
``a suitable site in which to define (integral) mixed Tate motives...''.
\par
May also pointed out that
his ``little convex bodies'' were described as a partial operad on page 172 of
\cite{M3},  but later was replaced by a strict operad by
Steiner \cite{Ste} in a way presaging Huang and Lepowsky's rescaling.
\vskip3ex
\Section {10. Batalin-Vilkovisky algebras}
\pp
Since punctured Riemann spheres or restricted polyhedra
(or, up to homotopy equivalence, configurations of points in
the plane) govern the structure of Gerstenhaber algebras,  what then is
the import of the `decorations' which correspond to torus bundles over
the component undecorated moduli spaces of the operads?
At the homology level, Getzler and Jones see the answer as the structure of
a Batalin-Vilkovisky algebra.
\pp
{\bf Definition:}
A {\bf Batalin-Vilkovisky algebra} is a  graded commutative algebra
with an operator $\Delta: A_{\bull}\to A_{\bull +1} $ such that
$\Delta^2 = 0$ and
$$
\align
\Delta(abc) &= \Delta(ab)c + (-1)^{|a|} a\Delta(bc)
+ (-1)^{(|a|-1)|b|} b\Delta(ac) \\
&- (\Delta a)bc - (-1)^{|a|}a(\Delta b)c - (-1)^{|a|+|b|} ab(\Delta c) .
\endalign
$$
\pp

The defining  condition describes
$\Delta$ as a `second order derivation', that is,
the deviation of $\Delta$ from being a derivation is in turn a derivation
\cite{PS}.  In fact, Penkava and Schwarz point out that in the super-algebra
context a Batalin-Vilkovisky algebra is the same thing as a super-commutative
associative algebra with an odd second order derivation
$\Delta:A_\bull@>>>A_{\bull+1}$ such that $\Delta^2=0$.
They prefer the alternate:
\pp
{\bf Definition:}
A {\bf Batalin-Vilkovisky algebra} is a Gerstenhaber
algebra with
operator $\Delta:A_\bull@>>>A_{\bull+1}$, such that $\Delta^2=0$, and
with product, bracket and $\Delta$ related by the formula
$$
[a,b] = (-1)^{|a|} \Delta(ab) - (-1)^{|a|} (\Delta a)b - a(\Delta b) .
$$
Furthermore, in a Batalin-Vilkovisky algebra, $\Delta$ satisfies the
formula
$$
\Delta[a,b] = [\Delta a,b] + (-1)^{|a|-1} [a,\Delta b] .
$$
%\end{theorem}
\vskip3ex
\Section{11. The Batalin-Vilkovisky formalism}
\pp
Why were physicists (Batalin and Vilkovisky) \cite {BaVi}
interested in such structures?
The answer is that they were interested in the Lagrangian approach to
field theories. (For a thorough treatment which is very accessible
mathematically but written by physicists, see the book by Henneaux and
Teitelboim \cite {HT}.)
This means that they were concerned with a variational
problem, finding the critical points of a functional $S$ defined on some
space of fields, most frequently functions on the jet bundle over some
mapping space.  If a classical Lagrangian `action' $S$ is invariant
with respect to `gauge symmetries'  (e.g.  diffeomorphisms of the space of
maps in question), there results a perturbative expansion
$$
S_t = \Sigma t^iS_i
$$
where $S_0$ is the original $S$.
Batalin and Vilkovisky enlarge the original space of field functionals
to a differential
graded algebra with an `anti-bracket' $(\ ,\ )$ such that the desired
$S_t$ is a solution of the equation
$(S_t, S_t) = 0,$
in complete analogy with the usual integrability equation in algebraic
deformation theory.
Indeed, after the fact, we can recognize the classical Batalin-Vilkovisky
algebra as a Gerstenhaber algebra.  However, just as in the
Batalin-Fradkin-Vilkovisky approach to constrained Hamiltonian problems
\cite {BFV}, the motivation was
not from the classical situation, but rather from the desire to quantize it.
To do so, they introduce a new operator $\Delta$ and the B-V
`master equation' becomes
$$
(S_t, S_t) = -i\hbar \Delta S_t.
$$
It is this structure that Getzler \cite{Get}, Getzler and Jones \cite{GJ} and
Penkava and Scwarz \cite{PS} formalize in their definition of
a B-V algebra.
\par
Although they derive this structure from an action of (the homology of) any
of the operads homotopy equivalent to ${\cal P}$,
the master equation solutions of Zwiebach's closed string field theory involve
the moduli spaces of Riemann surfaces
of genus $g$ with $k+1$ punctures.  Moreover,
the notion of an operad needs to be extended to allow sewing of two local
coordinates on the same Riemann surface (of genus $g$, producing a surface of
genus $g+1$); I am unaware of any known mathematical generalization of operad
which handles this structure.

\vskip3ex
\Section {12. Closed string field theory: Reprise with inner product}
\pp
Now that we are back to closed string field theory, we should notice that
the relevant action $S_t$ involves terms of the form
$$
< \phi_0\phi_1\dots\phi_N > =  < \phi_0\vert\ [\phi_1\dots\phi_N ]>
$$
which can be regarded as   fundamental
or as determined by the $N$-ary brackets via the inner product
$<\ \vert \ >$.
The fundamental inner product $<\phi_o\vert \phi_1>$ involves integration
over $S^1$ as well as a non-trivial inner product on the Hilbert space
$\H$ which is the space of fields.
\par
Thus we have an example of an {\bf sh Lie algebra with inner product} as in
Kontsevich's graph cohomology \cite {Ko}.  A   corresponding {\bf
sh associative algebra
with inner product} appears in the open string field theory of the Kyoto
group \cite{HIKKO}, although there
$$
< \phi_0\phi_1\phi_2\phi_3\phi_4 > = 0
$$
for reasons explained in \cite{S1}.
\vskip2ex
One of the subtleties of graph
cohomology is getting the signs right!  For the associahedra, the geometry is
obvious but the combinatorics of the signs were originally a bear!  Kontsevich
gets around this very cleverly by having an orientation of the vector space
${\Bbb R}^E\oplus  H^1  $
to handle the signs (where $E$ is the number of edges in the graph
and $H^1$ is that cohomology of the graph).
\vskip3ex
\Section {13. `Standard' Constructions and moduli space compactifications}
\pp
We have seen higher homotopy analogs of strict (differential graded) algebras,
both Lie and associative.  Just as sh Lie algebras can be
defined in terms of a coderivation on    $\bw sV$,
so an sh associative  algebra can be defined in terms of a coderivation
on the tensor coalgebra on the suspension of the vector space.
But what are the analogs for other structures, for example,
Gerstenhaber algebras?
\pp
\pc
At the time of the conference, I speculated that the
Lian-Zuckerman homotopies were  part of an action
of an operad for a strong homotopy analog of $\cal G$, i.e. that
there exist approriate higher homotopies of all orders.
Since the above descriptions
of $\cal L$ and $\cal G$ differ in the use of part versus all of the
homology of configuration space, it is to be hoped that the same will
be true for their sh analogs in using part versus all of the cells of
${\cal P}_N.$ (This is work in progress by the Penn-Princeton
string quartet: Huang, Kimura, Stasheff and Voronov.)\}
\pp
One way to define strong homotopy analogs of algebraic structure
controlled by an operad is to define actions of the operad {\it
up to strong homotopy} which in turn means that the
sequence of $\SS_k$-equivariant maps
$\theta_k:\OP(k)\times X^k@>>>X$
satisfy the appropriate compatibility conditions {\it up to strong
homotopy}.  This is a reasonably effective procedure since the
higher homotopy conditions are essentially cubical in their
combinatorics.  This approach follows that of
Sugawara \cite{Su} who defined strong homotopy maps of associative
spaces.  Lada \cite{Lad} and a later work of Hinich and Schechtman
\cite {HS1} provide details.
\pp
I did not resort to this definition for sh Lie algebras, believing that
use of the `standard' construction is conceptually and computationally
more useful becuse of its (comparatively) small size.
What then of a `standard construction' approach to BV algebras?
\pp
Answers of one sort are provided by Getzler and Jones \cite{GJ}  and
in an alternate form by Ginzburg and Kapranov \cite {GK}.
Ginzburg and Kapranov are concerned primarily with a special class of
quadratically related operads, called Koszul, which includes all of the
examples of interest in this talk.  They introduce the notion of the cobar
construction on an operad, which strictly speaking involves a linear dual,
whereas Getzler and Jones deal somewhat more naturally with the bar
construction, at the expense of introducing the dual notion of a co-operad.
\pc Both `bar' and `cobar' are slightly misleading as the constructions
make crucial use of the symmetric group actions and so are really closer
to the standard constructions used on Lie and commutative algebras.\}
\par
A {\bf co-operad} in a symmetric monoidal category $\Cal C$ is an operad in the
opposite symmetric monoidal category $\Cat^\op$. For
example, if $\OP$ is an operad in the category of chain complexes,
then the linear dual $\OP^*$ is a co-operad in the
category of cochain complexes.

If $\cal O$ is an operad in the category of chain complexes such that
$\cal O(1)=\Bbb C$, then, ``inspired by constructions of Boardman and Vogt
\cite{BoVo} and Ginzburg and Kapranov \cite{GK}'',
Getzler and Jones define the {\bf bar co-operad} $\BC\a$ of $\a$. As graded
vector space, it is essentially
$$
\BC\a(k) =  \sum_{T\in\Cal T(k)} \otimes_{v\in T} \ s\a(Inp(v))
$$
where $\Cal T (k)$ is the set of (isomorphism classes) of trees with $k$
numbered inputs and $v$ denotes
an internal vertex of $T$ while $Inp( v)$ denotes the set  of incoming edges
to $v$, trees being directed so there is only one outgoing edge at any
vertex.  In order to describe
the action of the permutation group $\SS_k$ on $\BC\a$,
they (like Ginzburg-Kapranov) have changed the indexing from integers
to finite sets.  The differential is induced from the operation on trees
which contracts an internal edge to a point, thus identifying its two
vertices.  The tricky question of signs is handled in \cite {GK} by
consistent use of the top exterior power of the vector space spanned
by the internal edges of a tree.
\pp
Similarly, there is a {\bf cobar operad} construction applicable to
any co-operad and bar and cobar stand in the usual adjoint relation
up to homotopy.  Ginzburg and Kapranov are concerned with such
adjoint relations or `duality', in particular with that between Lie
algebras and commutative associative algebras.  For any operad $\a$,
there is a notion of a strong homotopy object  over the `dual' operad
$\a^*$ as precisely a strict object over the graded linear dual
of $B\a$ (assuming some appropriate finite dimensionality).
The Hinich-Schechtman operad quasi-isomorphic to $\cal L$
can be described as the graded linear dual of $B\cal N$.  Since
$\cal N (k)$ is just a singleton, $B\cal N$ is described entirely
in terms of trees.
\pp
It is this description in terms of trees that provides the link with
the compactifications of moduli spaces or configuration spaces in
the work of Beilinson-Ginzburg \cite {BG1,2}.  In particular, for
Riemann surfaces of genus $0$ with $N+1$ labelled punctures, the appropriate
compactification is stratified with strata indexed by isomorphism
classes of rooted trees with $N$ labelled inputs.
\pc These compactifications provide operads as do Zwiebach's compact
``cut-off'' subspaces.\}
\vskip3ex
\Section {14. When there is no internal differential}
\pp
Although the cohomology associated to a topological
chiral algebra inherits a strict B-V algebra
structure for which higher homotopies are not needed,
there may still exist a sequence of higher order operations
on the cohomology.  These may derive from higher homotopies
in terms of the fields (before passing to cohomology); this
is what occurs in Zwiebach\cite Z .  Alternatively, these
may occur even if the fields form a strict differential graded
algebra with no terms of higher order.  The latter is familiar
in algebraic topology in the context
of Massey products or their H-space analogs (in the associative case). For
example, the linking of two circles may be detected by the cohomology
product of the complement of the link, but the non-trivial linking
of  three circles, no two of which are linked, in the configuration
known as the Borromean rings can be detected by a tri-linear Massey product.
\par
How can we make sense out of an sh Lie algebra   on the (co)homology level,
where there is no apparent differential $d_1$?
We can consider it to be $0$  and then interpret the relations $(J_N)$
accordingly.
We still have left a
sequence of compatibility conditions for the brackets of all orders, beginning
with the Jacobi identity for the bi-linear bracket.  It is this version that
occurs in the `homotopy Lie algebra' of Witten-Zwiebach \cite{WZ}, although it
is
expressed in the form
$$
0 = \{V,V\}
$$
where $V$ is a `vector field' acting as a derivation $\{V,\ \}$ corresponding
precisely to our $D$ with $d_1 =0.$  Moreover, Witten and Zwiebach write
their vector field in terms of a basis of ghosts as:
$$
c_{bc}^a\eta^b\eta^c\partial_a + c_{bcd}^a\eta^b\eta^c\eta^d\partial_a +\cdots
$$
where $\partial_a =\partial/\partial\eta^a$.
The term
$ c_{bc}^a\eta^b\eta^c\partial_a $
is a basis dependent way of writing the usual Lie algebra co-boundary $d_2$
and the terms of higher order correspond precisely to the further $d_i.$
\vskip2ex
One explanation of such higher order structure is given by
the characteristic technique of Homological Perturbation
Theory (HPT) \cite{HPT}:
When    a chain complex (a differential graded module) $C$ has the structure
of a strict algebra, higher order operations in homology can arise as follows.
Assume we are over a ground field or that the  modules of concern
are free over a ground ring.  Suppose that
a chain complex (a differential graded module) $C$ has the structure
of a strict algebra.
Choose a splitting $C = H\oplus X$ where
$X$ is a contractible chain complex; indeed, choose a contracting homotopy.
HPT then provides an algorithm for constructing suitably compatible higher
order
 operations on the (co)homology $H$.
\vskip2ex
In the ``harmonic'' case , there is a particularly  good Hodge decomposition,
a
choice of splitting $H\hookrightarrow C$ which is a strict map of the
structures.    In physical language, Zwiebach says this as: ``the product of
physical states can not give an unphysical state'', meaning, at the chain
level,
that $H$ as a subspace of $C$ is closed under all the $N$-ary operations.
\vskip2ex
Finally let me call attention to a potential problem with terminology, namely,
 a conflict with the terminology in algebraic topology if
`homotopy Lie algebra' is used to indicate not only the first order
homotopy for the Jacobi identity, but also the higher homotopies which
I indicate by `strong' or `strongly'.
\vfill
\eject

%\input amstex
%\magnification=\magstep1
 \hsize=15cm
 \vsize=19.5cm
\def\parag{\vskip3ex}
 \pageno=20
 \headline={\ifnum\pageno>1 \tenbf
 Closed SFT and operads      \hss Stasheff\else\hfil\fi}
\documentstyle{amsppt}
\widestnumber\key{HIKKO}
\NoBlackBoxes
%\magnification=\magstep2
\def\cal{\Cal}
\def\fp{\flushpar}
\def\no{\key}
\def\inseries{\bookinfo}


\Refs
\ref\key  Ad\by J. F. Adams
\book Infinite loop spaces
\publ Princeton U Press, Princeton NJ \yr 1978
\endref
\fp

\ref\no Ar\by   V.I. Arnol'd
\paper The cohomology ring of the colored braid group
\jour Mat.  Zametki  \yr 1969  \pages 227--231
\endref

\fp
\ref\no BF\by       I.A. Batalin and E.S. Fradkin
\paper A generalized canonical formalism and quantization of reducible gauge
theories
 \jour Phys. Lett. \vol 122B \yr (1983) \pages 157-164
\endref
\fp
\ref\no BaVi \by I.A. Batalin and G.S. Vilkovisky
\paper Existence theorem for gauge algebra \jour J. Math. Phys.
\vol 26 \yr (1985) \pages 172-184
\endref
\fp
\ref\key  \by  I.A. Batalin and G.S. Vilkovisky
\paper Quantization of gauge theories with linearly dependent generators
\jour  Phys. Rev. D \vol 28 \yr (1983) \pages 2567-2582
\endref
\fp
\ref\no    \by   I.A. Batalin and G.S. Vilkovisky
\paper Relativistic S-matrix of dynamical systems with boson and fermion
constraints
\jour Phys. Lett. \vol 69B \yr (1977) \pages 309-312
\endref
\fp
\ref\no{BG1}\by    A.~Beilinson and V.~Ginzburg
\paper Infinitesimal structures of moduli space of $g$-bundles
\jour Duke Math. J. \vol  66 \yr 1992 \pages 63-74
\endref
\fp
\ref\no BG2 \quad \by A. Beilinson and V. Ginzburg
\paper Resolution of diagonals, homotopy algebra and moduli spaces
\jour preprint    \yr  \vol  \pages
\endref
\fp
\ref \no {BBvD} \by F.A. Berends, G.J.H. Burgers and H. van Dam
\paper On the theoretical problems in constructing interactions involving
higher spin massless particles
\jour  preprint IFP234-UNC \yr 1984
\endref
\fp
\ref\no \hskip8ex
\by G.J.H. Burgers
\paper On the construction of field theories for higher spin
                massless particles
\paperinfo doctoral dissertation
\jour Rijksuniversiteit te Leiden  \yr 1985
\endref
\fp
\ref\no {BoVo}\by J.M. Boardman and R.M. Vogt
\book  Homotopy invariant algebraic structures on topological spaces
\inseries Lecture Notes in Mathematics
\vol 347 \publ Springer-Verlag, Berlin-New York \yr 1973
\endref
\fp

\ref \no CE \by C.~Chevalley and S.~Eilenberg
\paper  Cohomology theory of Lie groups and Lie algebras
\jour Trans. Amer. Math. Soc.
\vol 63 \pages 85-124
\endref
\fp
\ref\no {C1}\by F.R. Cohen
\paper   The homology of {${\cal C}_{n+1}$}-spaces, {$n\ge0$}
\inbook      The homology of iterated loop spaces
\inseries Lecture Notes in Mathematics\vol  533
\publ Springer-Verlag, Berlin-New York
\yr  1976 \pages 207--351
\endref
\fp
\ref \no {C2} \by F. R. Cohen
\paper
Artin's braid groups, classical homotopy theory and sundry other curiosities
\jour Contemp. Math. \vol 78 \yr 1988 \pages 167-206
\endref
\fp
\ref\no D \by P. Deligne\paper  Th\'eorie de Hodge II
\jour  Publ. IHES\vol  40\yr  1971\pages 5-58
\endref
\fp
\ref\no{FM} \by W.  Fulton and  R.  MacPherson
\paper A compactification of configuration space
\jour preprint
\endref
\fp
\ref\no {Ger}\by M. Gerstenhaber
\paper The cohomology structure of an associative ring
\jour Ann. of Math. \vol 78\yr 1962 \pages 267-288
\endref
\fp
\ref\no  \by M. Gerstenhaber
\paper On the deformation of rings and algebras
\jour Ann. of Math. \vol 79\yr 1964\pages  59-103
\endref
\fp
\ref\no {Get}\by E. Getzler
\paper Batalin-Vilkovisky algebras and two-dimmensional topological field
theories
\publ preprint
\endref
\fp
\ref\no {GJ} \by E. Getzler and J.D.S. Jones
\paper n-algebras and Batalin-Vilkovisky algebras
\publ preprint
\endref
\fp
\ref\no {GK} \by V. Ginzburg and M. Kapranov
\paper Koszul duality for operads
\publ preprint
\endref
\fp
\ref \no HIKKO \by H. Hata, K. Itoh, T. Kugo, H. Kunitomo and K. Ogawa
\paper Covariant string field theory
\jour Physical Review D \vol 34 \yr 1986 \pages 2360-2429
\endref
\fp
\ref \by  H. Hata, K. Itoh, T. Kugo, H. Kunitomo and K. Ogawa
\paper Covariant string field theory II
\jour  Physical Review D \vol 35 \yr 1987
\pages 1318-1355
\endref
\fp
\ref \by  T. Kugo
\paper String field theory
\paperinfo Lectures delivered at 25th
             Course of the International School of Subnuclear Physics on ``The
             SuperWorld II'', Erice, August 6-14, 1987
\endref
\ref\no HT \by M. Henneaux and C. Teitelboim
\book Quantization of Gauge Systems
\publ Princeton Univ. Press \yr 1991
\endref
\fp
\ref \no HS1 \by V. Hinich and V.Schechtman
\paper On the homotopy limit of homotopy algebras
\jour LNM \vol 1289 \yr 1987 \pages 240-264
\publ Springer-Verlag
\endref
\fp
\ref\no{HS2} \by V. Hinich and V. Schechtman
\paper Homotopy Lie algebras
\jour preprint
\endref
\fp
\ref\no {HKR} \by G. Hochschild, B. Kostant and A, Rosenberg
\paper Differential forms on regular affine algebras
\jour Trans AMS \vol 102 \yr 1962 \pages 383-408
\endref
\fp
\ref \no HPT \by D. Barnes and L.A. Lambe
\paper A fixed point approach to homological perturbation theory
\jour Proc. AMS \vol 112   \yr 1991   \pages 881-892
\endref
\fp

\ref \by V.K.A.M. Gugenheim
\paper  On the chain complex of a fibration
\jour Ill. J. Math. \vol 3 \yr 1972 \pages 398-414
\endref

\ref
\by V.K.A.M.  Gugenheim
\paper On a perturbation theory for the homology of the loop-space
\jour J. Pure \& Appl. Alg.\vol 25 \yr 1982 \pages 197-205
\endref
\ref
\by V.K.A.M.  Gugenheim and L. Lambe
\paper Applications of perturbation theory in differential homological algebra
I
\jour Ill. J. Math.\vol 33 \yr 1989 \pages 566-582
\endref

\ref
\by V.K.A.M.  Gugenheim, L. Lambe and J. Stasheff
\paper  Algebraic aspects of Chen's twisting cochain
\jour Ill. J. Math \vol 34 \yr 1990 \pages 485-502
\endref

\ref
\by V.K.A.M.  Gugenheim, L. Lambe and J. Stasheff
\paper Perturbation theory in differential homological algebra II
\jour Ill. J. Math.\vol 35   \yr 1991   \pages     357-373
\endref
\ref
\by V.K.A.M.  Gugenheim   and J. Stasheff
\paper On perturbations and $A_\infty$-structures
\jour Bull. Soc. Math. de Belg.
\vol  38 \yr 1986 \pages 237-246
\endref

\ref \by J. Huebschmann
\paper Perturbation thoery and small methods for the chains of certain
induced fibre spaces
\jour Habilitationsschrift \publ Universit\"at Heidelberg \yr 1984 \endref

\ref \by J. Huebschmann \paper
The cohomology of $F\Psi^q$. The additive structure \jour J. Pure and Appl.
Alg. \vol 45 \yr 1987 \pages 73-91\endref

\ref \by J. Huebschmann \paper
Perturbation theory and free resolutions for nilpotent groups of class 2
\jour J. Alg. \vol 126 \yr 1989 \pages 348-399 \endref


\ref \by J. Huebschmann \paper
Minimal free multi models for chain algebras \jour preprint \yr 1990
\endref

\ref \by J. Huebschmann and T. Kadeishvili
\paper Small models for chain algebras \jour Math. Z.
\vol 207 \yr 1991 \pages 245-280 \endref

\ref  \by L. Lambe
\paper  Homological Perturbation Theory - Hochschild Homology and Formal Groups
\inbook Proc. Conference on Deformation Theory and Quantization with
Applications to Physics, Amherst, June 1990
\inseries Contemporary Math \yr 1992      \publ AMS
\endref

\ref
\by L. Lambe and J.D. Stasheff
\paper  Applications of perturbation theory to iterated fibrations
\jour Manuscripta Math. \vol 58 \yr 1987 \pages 363-376
\endref

\ref\key Hu \by   Y.-Z. Huang
\paper         On the geometric interpretation of vertex operator algebras
\jour     Ph.D. thesis, Rutgers University \yr 1990
\endref
\ref \by Y.-Z. Huang
\paper    Operads and the geometric interpretation of vertex operator algebras,
I,
\jour preprint
\endref
\fp
\ref \by Y.-Z. Huang
%\bibitem[H2]{H2}
\paper    Geometric interpretation of vertex operator algebras,
\jour  Proc. Natl. Acad. Sci. USA \vol 88 \yr 1991 \pages 9964--9968
\endref
\fp
\ref \by Y.-Z. Huang
%\bibitem[H3]{H3}
\paper    Applications of the geometric interpretation of vertex operator
algebras,
\inbook   Proc. 20th International Conference on
Differential Geometric Methods in Theoretical Physics, New York, 1991
\eds      S. Catto and A. Rocha
\publ     World Scientific, Singapore \yr 1992 \vol 1 \pages 333--343
\endref
\fp
\ref \by Y.-Z. Huang
\paper    Operads and the geometric interpretation of vertex operator algebras,
II,
\jour     in preparation.
%this is the draft I used for USC
\endref
\fp
\ref\key HL \by Y.-Z. Huang and J. Lepowsky
\paper    Toward a theory of tensor products for representations of a vertex
operator algebra,
\inbook   Proc. 20th International Conference on Differential Geometric
Methods in Theoretical Physics, New York, 1991
\eds      S. Catto and A. Rocha
\publ     World Scientific, Singapore \yr 1992 \vol 1  \pages 344--354
\endref
\fp
\ref
\by Y.-Z. Huang and J. Lepowsky
\paper    Vertex operator algebras and operads
\book The Gelfand Mathematical Seminars, 1990--1992
\ed  L. Corwin,  I. Gelfand and J. Lepowsky
\publ Birkh\"{a}user, Boston \yr 1993  \pages 145-161
\endref
\fp
\ref \by Yi-Zhi Huang and James Lepowsky
 \paper  Operadic formulation of the notion of vertex operator algebra
\endref
\fp

\ref \no K \by M. Kaku
\paper Why are there two BRST string field theories?
\jour Phys. Lett. B
\vol 200 \yr 1988 \pages 22-30
\endref
\fp
\ref \by M. Kaku
\paper Deriving the four-string interaction from geometric string field theory
\jour preprint, CCNY-HEP-88/5
\endref
\fp
\ref \by M. Kaku
\paper Geometric derivation of string field theory from first
             principles: Closed strings and modular invariance
\jour preprint, CCNY-HEP-88/6
\endref
\fp
\ref \by M. Kaku
\book Introduction to Superstrings
\publ Springer-Verlag \yr 1988
\endref
\fp
\ref \by M.Kaku and J. Lykken
\paper Modular invariant closed string field theory
\jour preprint, CCNY-HEP-88/7
\endref
\fp
\ref\no{Ke}\by M. Kelly
\paper
\jour preprint
\endref
\fp
\ref\no {Ko}\by M. Kontsevich
\paper Graphs, homotopical algebra and low-dimensional topology
\publ preprint
\endref
\fp
\ref\no \by M. Kontsevich
\paper  Feyneman diagrams and low-dimensional topology
\publ preprint (to appear in
\book
\endref
\fp
\ref\no Kn \by F.F.  Knudsen
\paper The projectivity of the moduli space of
stable curves II. The stacks $\overline {M_{g,n}}$
\jour  Math.  Scand.\vol 52 \yr 1983\pages 161 - 189
\endref
\fp
\ref\no  {KM} \by I. Kriz and J. P. May
\paper DGA's up to homotopy and their derived categories
\publ preprint, being revised as
\endref
\fp
\ref\no   \by I. Kriz and J. P. May
\paper Operads, algebras and modules
\publ preprint
\endref
\fp
\ref \no KKS \by T. Kugo, H. Kunitomo and K. Suehiro
\paper Non-polynomial closed string field theory
\jour  Phys. Lett. \vol 226B \yr 1989 \pages 48-54
\endref
\fp
\ref \by  T. Kugo and K. Suehiro
\paper Nonpolynomial closed string field theory: Action and gauge invariance
\jour Nucl. Phys. B \vol 337 \yr 1990 \pages 434-466
\endref
\fp
\ref\no {Lad}\by T. Lada
\paper Strong homotopy algebras over monads
\inbook      The homology of iterated loop spaces
\inseries Lecture Notes in Mathematics\vol  533
\publ Springer-Verlag, Berlin-New York
\yr  1976 \pages 207--351
\endref

\fp
\ref\no {LS} \by T. Lada and J. Stasheff
\paper Introduction to sh Lie algebras for physicists
\jour preprint HEP-TH 9209099 \endref
\fp
\ref\no{Las}\by R. Lashof
\paper Classification of fibre bundles by the loop space of the base
\jour Annals of Math. \vol 656 \pages  436-4
\endref
\fp
\ref\no{Law} \by W. Lawvere
\paper Some algebraic problems in the context of functorial semantics
of algebraic theories
\inbook Report of the MidWest Category Seminar II
\publ Springer-Verlag \inseries Lecture Notes in Mathematics \vol 61
%This paper contains citations to the earlier appearances.
\endref
\fp
\ref\no {LZ} \by B. Lian and G. Zuckerman
\paper New Perspectives on the  BRST-algebraic Structure of String Theory
\publ preprint
\endref
\fp
\ref\no {M1}\by J. P. May
\book The geometry of iterated loop spaces
\inseries Lecture Notes in Mathematics
\publ Springer-Verlag \vol  271 \yr 1972
\endref
\fp
\ref\no {M2}\by J. P. May
\paper Infinite loop space theory
\jour Bull. AMS \vol 83 \yr 197
456-494.
\endref
\fp
\ref\no {M3}\by J. P. May
\book  $E_{\infty}$-ring spaces and $E_{\infty}$-ring spectra
\publ Springer-Verlag
\inseries  LNM \vol 577 \yr 1977
\endref
\fp
\ref\no {PS}  \by M. Penkava and A. Schwarz
\paper  On some algebraic structures arising in string theory
\jour preprint UCD-92-03, HEP-TH/9212072
\endref
\fp
\ref\no {R}\by I. Rivin
\paper A characterization of ideal polyhedra in hyperbolic 3-space
\jour preprint Princeton 1992
\endref
\fp
 \ref \by C.D. Hodgson, I. Rivin and W.D. Smith
\paper A characterization of convex hyperbolic polyhedra  and of convex
polyhedra inscribed in the sphere
\jour Bull AMS \yr 1992 \vol \pages
\endref
\fp
\ref \no {SZ} \by M. Saadi and B. Zwiebach
\paper Closed string field theory from polyhedra
\jour Ann. Phys. (N.Y.) \vol 192 \pages 213-227\yr 1989
\endref
\fp
\ref \no {SS} \by M. Schlessinger and J.~D.~Stasheff
\paper The Lie algebra structure of tangent cohomology and
deformation theory
\jour J. of Pure and Appl. Alg. \vol 38   \yr 1985   \pages 313-322
\endref
\fp
\ref  \by M.~Schlessinger and J.~D.~Stasheff
\paper Deformation theory and rational homotopy type
\jour Publ. Math. IHES
\yr to appear - eventually
\endref
\fp
\ref \no S1 \by J.~Stasheff
\paper An almost groupoid structure for the space of (open) strings and
implications for string field theory
\inbook Advances in Homotopy Theory (Cortona, June 1988)
\inseries LMS Lecture Notes Series \vol 139 \yr 1989 \pages 165-172
\endref
\fp
\ref \no {S3} \by J.D.~Stasheff
\book  H-spaces from a homotopy point of view
\inseries Lecture Notes in Mathematics \vol 161
 \publ  Springer-Verlag
\yr 1970
\endref
\fp
\ref\no  {S4} \by  J.D.~Stasheff
\paper On the homotopy associativity of H-spaces I
\jour Trans. AMS
\vol 108\pages 275 -292
\yr 1963
\endref
\fp
\ref \no {S5} \by J.D.~Stasheff
\paper On the homotopy associativity of H-spaces II
\jour Trans. AMS \vol 108\pages 293-312
\yr 1963
\endref
\fp
\ref \no {S6} \by J.~Stasheff
\paper Constrained Poisson algebras and strong homotopy representations
\jour Bull. AMS \vol 19 \yr 1988 \pages 287-290
\endref
\fp
\ref
\by J. Stasheff
\paper Homological (ghost) approach to constrained Hamiltonian systems
\inbook Proceedings of the Conference on Mathematical Aspects of Classical
Field Theory, Seattle, July 1991
\inseries   Contemporary Mathematics Series \vol \yr \pages
\endref
\fp
\ref\no{Ste}\by R. Steiner
\paper A canonical operad pair
\jour Math Proc Camb Phil Soc \vol 85 \yr 1979\pages  443-449
\endref
\fp
\ref\no Su \by M. Sugawara
\paper On the homotopy-commutativity of groups and loop spaces
\jour Mem. Coll. Sci. Univ. Kyoto Ser. A Math.
\vol 33 \yr 1960/61 pages 257-269
\endref
\fp
\ref \no Wies \by H.-W.~Wiesbrock
\paper A note on the construction of the $C^*$-Algebra of bosonic strings
\jour JMP \vol 33 \pages 1837- \yr 1992
\endref
\fp
\ref  \by H.-W.~Wiesbrock
\paper The C*-Algebra of Bosonic Strings
\jour CMP \vol 136 \pages 369-397 \yr 1991
\endref
\fp
\ref  \by H.-W.~Wiesbrock
\paper The Construction of the sh-Lie-Algebra of Closed Bosonic Strings
\jour CMP \vol 145 \pages 17-42 \yr 1992
\endref
\fp
\ref  \by H.-W.~Wiesbrock
\paper The quantum algebra of bosonic strings
\jour preprint FUB-HEP\slash 89-9
\endref
\fp
\ref  \by H.-W.~Wiesbrock
\paper The mathematics of the string algebra
\jour preprint DESY 90-003
\endref
\fp
\ref \no W1 \by E.~Witten
\paper Non-commutative geometry and string field theory
\jour Nuclear Physics B \vol 268 \yr 1986 \pages 253-294
\endref
\fp
\ref \by E.~Witten
\paper Interacting field theory of open strings
\jour Nuclear Physics B \vol 276 \yr 1986 \pages 291-324
\endref
\fp
\fp \ref \no{W2}
%\bibitem{W2}
\by E. Witten\paper The anti-bracket formalism
\jour preprint IASSNS-HEP-90/9
\endref
\fp
\ref\key{WZ}
\by E. Witten and B. Zwiebach
\paper
Algebraic structures and differential geometry in two-dimensional
string theory
\jour Nucl. Phys. B\vol 377 \yr 1992 \pages 55-112
\endref
\fp
\ref\key {Z}\by B. Zwiebach
\paper Closed string field theory: quantum action and the B-V master equation
\jour Nucl. Phys. B\vol 390 \yr 1993 \pages 33-152
\endref
\endRefs
\enddocument